# Data filtering methods for SARS-CoV-2 wastewater surveillance

Rezgar Arabzadeh*, Daniel Martin Grünbacher*, Heribert Insam**, Norbert Kreuzinger***, Rudolf Markt** and Wolfgang Rauch*^

* Unit of Environmental Engineering, Department of Infrastructure, University of Innsbruck, Technikerstrasse 13, 6020 Innsbruck, Austria;
** Department of Microbiology, University of Innsbruck, Austria
*** Institute for Water Quality and Resource Management, Technische Universität Wien, Austria
^ Corresponding author        (E-mail: wolfgang.rauch@uibk.ac.at)

**Abstract:**

In the case of SARS-CoV-2 pandemic management, wastewater-based epidemiology aims to derive information on the infection dynamics by monitoring virus concentrations in the wastewater. However, due to the intrinsic random fluctuations of the viral signal in the wastewater (due to e.g., dilution; transport and fate processes in sewer system; variation in the number of persons discharging; variations in virus excretion and water consumption per day) the subsequent prevalence analysis may result in misleading conclusions. It is thus helpful to apply data filtering techniques to reduce the noise in the signal. In this paper we investigate 13 smoothing algorithms applied to the virus signals monitored in four wastewater treatment plants in Austria. The parameters of the algorithms have been defined by an optimization procedure aiming for performance metrics. The results are further investigated by means of a cluster analysis. While all algorithms are in principle applicable, SPLINE, Generalized Additive Model and Friedman's Super Smoother are recognized as superior methods in this context (with the latter two having a tendency to over-smoothing). A first analysis of the resulting datasets indicates the influence of catchment size for wastewater-based epidemiology as smaller communities both reveal a signal threshold before any relation with infection dynamics is visible and also a higher sensitivity towards infection clusters.

**Highlights**

- The timeline of SARS-CoV-2 virus concentration - monitored in four wastewater treatment plants across Austria over a duration of approx. 8 months – reveals a significant random component
- Multiple filtering techniques are investigated for their potential to smooth the virus signals
- Based on parameter optimization and cluster analysis SPLINE and Generalized Additive Model are seen as superior algorithms for smoothing Sars-Cov2 signals

**Introduction**

Management of the SARS-CoV-2 pandemic rests upon measures such as hygiene, isolation and vaccination but requires rigorous monitoring on the state and spread of the disease (Nicola et al. 2020). Next to individual qPCR and antigen testing, wastewater-based epidemiology (WBE) has been recognized as a valuable tool to estimate viral prevalence (Weidhaas et al., 2021). There is a rapidly increasing body of evidence about the methodology and its application (see e.g., Ahmed et al. 2020; Kitajima et al. 2020, Zhu et al., 2021). Several studies, e.g. Wölfel et al. (2020), showed that infected persons in a catchment shed a certain amount of viral load per day into the sewer system, resulting in measurable viral titer at the sampling point – expressed as virus RNA concentrations [number of RNA copies/ml]. Such measurements are usually taken as composite samples at the treatment plant of the urban drainage system.

Key element in WBE is to derive information on the infection dynamics in the catchment by means of the monitored virus particle concentrations by quantifying their RNA genome. The signal serves as a proxy for prevalence, i.e., the total number of infected persons in the catchment (Zheng et al. 2019). WBE is thus a valuable additional source of information next to individual testing strategies. Even more, time series prediction can be applied to the signal, thus serving as potential early-warning tool in pandemic management (e.g., Hart et al. 2020; Gonzalez et al. 2020).

Relating to the basic concepts of WBE (Feng et al., 2018) it is typical not only to use the raw signal (RNA concentration as equivalent to virus particles) for further analysis but to apply normalization regarding a) flow dynamics and b) changes in population by use of biomarkers (Tscharke et al., 2019; Been et al., 2014). Still, it is due to the complexity of the whole process that the wastewater titer signals (both raw value and normalized) not only express the prevalence information but also contain huge variations. Reasons are manifold, but key factors are a) the individual variances in viral shedding (both amount and time) of the infected persons b) effect of spatial distribution of the viral load in the catchment (i.e., the location of the main entry points) c) stochastic influences to transport mechanics and virus degradation in the sewer system d) influence of rain runoff due to dilution and loss via CSO and e) variances that stem from both the sampling procedure and the laboratory methods. To conclude, the wastewater signal (even if normalized) contains not only the sought-after information on prevalence in the catchment but also a large noise contribution. The latter, however, makes the analysis and data interpretation difficult and data modeling a poorer fit to the actual status (Wand 2003).

To differentiate the noise from the actual information in the signal, filtering techniques are frequently used in science and engineering (Huang et al. 2016). Simple filter techniques such as moving average are common, but recent studies (e.g., Stadler et al. (2020) and Nemudryi et al. (2020)) applied more advanced procedures such as locally weighted polynomial or spline methods to smooth the viral loads signals. However, a thorough investigation to identify the optimal smoothing method applicable to filter SARS-Cov-2 time series is still missing. In this study, 13 filtering methods – from simple to advanced - have been applied to the viral load measured at four locations in Austria and are investigated towards three performance indicators, i.e., mean absolute error (MAE), variability (VAR), and Akaike Information Criterion (AIK).

In the remainder of the paper, we first outline the status of WBE in Austria, the selected four case studies and the titer datasets derived therefrom. The filtering methods are presented, however not elaborated in detail. To rationalize optimal performance, an optimization procedure is used for performance metric. The results are further investigated by means of a cluster analysis. Last, the explanatory power of the datasets with respect to infection dynamics is analyzed.

**Materials and Methods**

*Wastewater surveillance and datasets*

Already early in the pandemic Austria has established the research project Coron-A to develop the scientific background of WBE as Covid-19 surveillance tool (Coron-A, 2021). Fundamental in the project is the surveillance of up to 44 wastewater treatment plants (WWTPs) in Austria by taking (at least) bi-weekly 24h composite volume proportional samples (CVVT: constant volume; variable time) from the inlet of the WWTPs.

Sampling is done by cooled automatic samplers (various suppliers). Samples are cooled to 4°C (Markt et al., 2021) and shipped to the laboratory in Styrofoam boxes with coolpacks guaranteeing continuous cooling during transport. In the laboratory, samples of 50-100g are centrifuged for 30 min at 4,500 g (4508 R cooling centrifuge, Eppendorf, Hamburg, Germany) (Medema et al. 2020) to remove particulate matter. The supernatant is then concentrated through polyethylene glycol (PEG) centrifugation at 12,000 g for 99 min. The pellet obtained is suspended in 800-1000 µl lysis buffer (details see Markt et al., 2021) and transferred to a micro reaction tube (Eppendorf). The RNA is purified using Monarch™ total RNA Miniprep Kit (New EnglandBiolabs, Ipswich, USA). After Nanodrop RNA quantification and appropriate dilution the SARS-CoV-2 nucleocapsid (N1) gene RNA copy numbers are determined on a RotorGene cycler (Qiagen, Hilden, Germany) using a plasmid standard containing the N-gene of Sars-COV-2 (2019-nCoV_N_Positive Control, IDT, Leuven, Belgium) (Markt et al., 2021).

According to national regulations, WWTP's in Austria apply a self-monitoring scheme and measure flow rate and temperature on a daily basis. Water quality parameters such as COD, $N_{tot}$ and $NH_4^+$ are analyzed as well, but the frequency depends on the design capacity of the investigated WWTP (varying between daily to weekly). Water quality parameters are likewise determined via the same 24h composite samples as used for determination of the SARS-CoV-2 titer.

For our study, we selected four different sampling locations or urban drainage catchments respectively (see Table 1). For easier reference and respecting data protection acts in Austria, the catchments/cities are denoted as A-D in the following. The locations mainly vary in population and catchment area size as well as type of sewage system.

Table 1: Sampling locations – served population and climatic conditions

| Sampling site | | Connected residents | Avg. daily inflow 2020 (m³/d) | Avg. monthly temperature 1971-2000 (°C) * | Avg. total annual precipitation 1971-2000 (mm/a) * | Avg. number of days with total daily precipitation > 10 mm (d/a) * |
|---|---|---|---|---|---|---|
| Urban | A | 1900000 | 539500 | 11.4 | 548 | 14.9 |
| | B | 320700 | 83190 | 9.0 | 1184 | 40.0 |
| Rural | C | 41700 | 16340 | 8.9 | 1231 | 40.3 |
| | D | 23600 | 4900 | 7.9 | 889 | 29.1 |

* Zentralanstalt für Meteorologie und Geodynamik ZAMG (2002)

Case studies A and B represent prototypical Austrian cities (large to medium) with high population density and an urban environment. In both cases, the entity of the urban catchment is discharged to the WWTP. Case studies C and D, on the other hand, resemble smaller settlements and case study D is moreover a highly touristic place with predominately summer tourism. Meteorological data from 1971 to 2000 shows a temperate climate for all sampling sites. However, the locations experience up to 40 days/year with a total daily precipitation of 10 mm or higher which leads to significant runoff and to a loss of virus particles in the sewage by combined sewer overflow.

Wastewater surveillance in the 4 chosen case studies started in summer 2020 (in case of location A already in May 2020) and samples were taken weekly or more frequently. In this study we concern ourselves with the data until end of 2020, that is a timeline of 8 months – see Figure 1 for details.

The timeline of the (raw) wastewater titer values follows the epidemic data (expressed here as active cases as identified by means of PCR tests) in Austria (Figure 1). The lockdown after the first pandemic wave in March 2020 was quite successful in reducing also the virus signal in wastewater. This is demonstrated by the low RNA concentration measured at city A in the early summer period. In principle, the signal remained at a low level for all sites during summer and early fall. One exception was case study D where a sharp increase in RNA concentrations was observed in August, matching local infection dynamics caused by tourism. Starting with October 2020 the beginning of the second Austrian pandemic wave was depicted also in the wastewater signal. Both the reported cases of infections and the viral RNA concentration in the 4 WWTPs peaked in mid-November. Thereafter, another lockdown has been imposed over the country that again declined both the infections and the RNA signal.

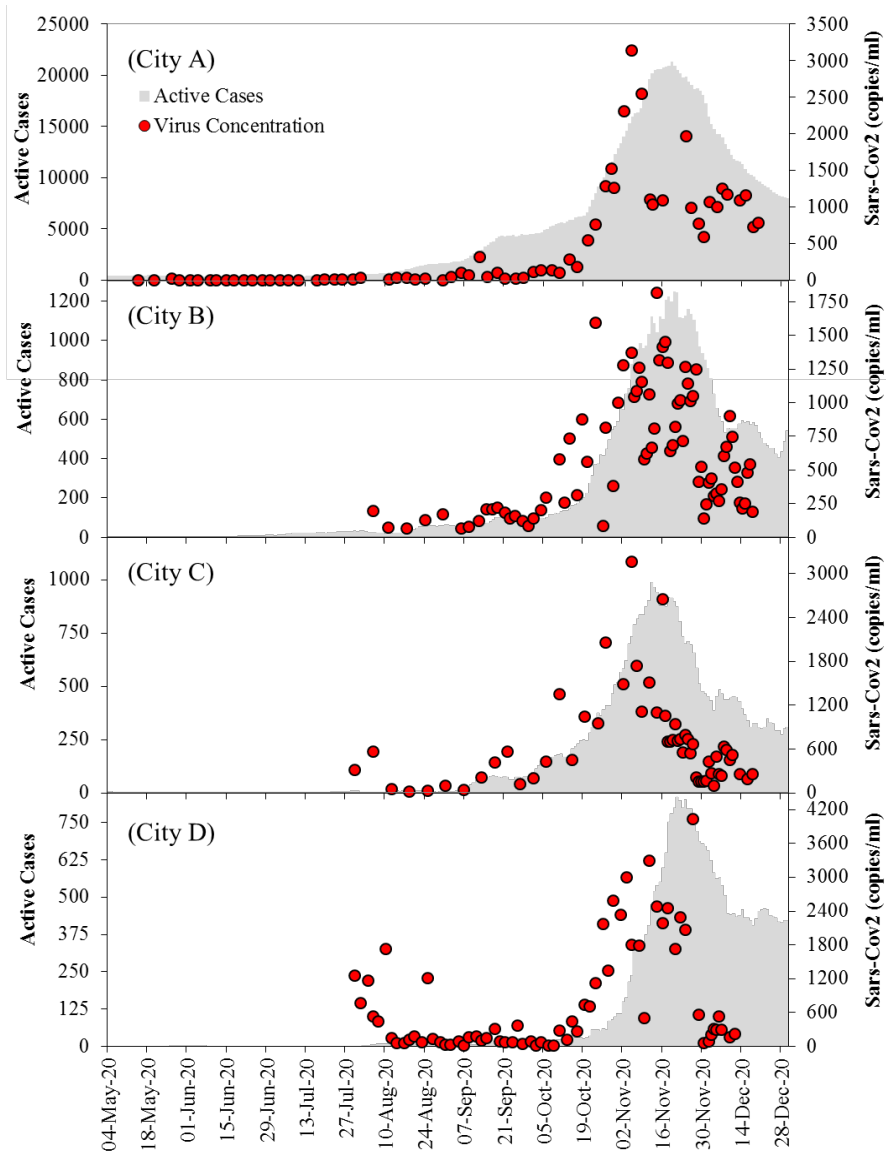

Figure 1: Raw data timelines of SARS-CoV-2 titer values (RNA copies/ml) and epidemiological timelines (active cases as identified by PCR-tests) at the 4 sampling sites.

*Normalization*

Wastewater-based epidemiology for pandemic management aims at deriving information on prevalence in the catchment. Prevalence is here defined as the fraction of the infected persons within the total population discharging into the sewer system. If – for the sake of simplicity - we assume that each infected person sheds a certain virus load per day, we can relate the measured virus concentration $c_{virus}$ to the infection dynamics. We are thus less interested in the raw surveillance data but in the specific viral load instead and derive - for an arbitrary datapoint in the series:

$$L_{virus} = \frac{c_{virus} * Q}{P} \qquad \text{Eq.1}$$

Where $L_{virus}$ is RNA copies/P/d; $Q$ = flow volume in L/d; $c_{virus}$ = virus concentration in the sample in RNA copies/L and P = number of persons in the watershed. While the consideration of flow in the timeline is evident from the measured inflow data at the WWTP (see Table 1) the temporal variation of P in the catchment is to be estimated via a wastewater biomarker (Chen et al, 2014) as:

$$P = \frac{c_{bm}*Q}{f_{bm}} \qquad \text{Eq.2}$$

Where $c_{bm}$ is the concentration of biomarker in g/L and $f_{bm}$ = specific biomarker load in g/P/d. The choice of an appropriate biomarker has been subject to numerous investigations (Choi et al., 2018). However, for this investigation we are less interested in actual values but can express the influence by normalization. As biomarker readily available at wastewater treatment plants we apply the standard water quality parameter $NH_4$-N and derive the specific load $f_{NH4}$ from the measured 50-percentile value in the period of the first lockdown in Austria as load fluctuations are minimal therein (Table 2). The normalized signal is calculated as

$$L_{virus} = \frac{c_{virus}*f_{NH4}}{c_{NH4}} \text{ in RNA copies/cap/day} \qquad \text{Eq.3}$$

Table 2: Specific $NH_4$-N load in g per capita per day. Calculated percentiles from daily measurements during the first lockdown period in Austria (April to mid-May 2020)

|  | City A | City B | City C | City D |
| --- | --- | --- | --- | --- |
| 2.5% percentile | 9.77 | 5.84 | 8.02 | 5.94 |
| 50.0% percentile | 10.71 | 6.49 | 8.99 | 6.80 |
| 97.5% percentile | 12.17 | 7.13 | 9.73 | 9.32 |

*Filtering techniques*

The timeline of surveillance raw data (and normalized data as well) includes not only the sought-after information regarding prevalence in the catchment but contains a significant noise contribution, that is due to stochastic effects in the whole process. In this study we apply and compare 13 filter/smoothing techniques with the aim to de-noise the time series of RNA-concentration in WBE. Table 3 summarizes the methods applied herein and gives the key reference(s) for each.

Typically, the filtering methods can be discriminated by the parameters needed for its use, ranging between 0 and more than 3. Parameter-less data models (here FFT, TUK, and KAF) are mathematically complex and (usually) computationally more difficult to implement, the benefit being obvious as calibration is omitted for these techniques. Data models typically have one to three adjustable parameters. The most common used parametric method in the engineering community is (centralized) SMA which is both simple to implement and contains one parameter only. The detriment is the lower robustness as compared to other – more complex – methods (Raudys et al. 2018). The GAM model, which is a more recent development, contains a high number of parameters and is a combination of additive models and generalized linear ones (Wood et al., 2016). However, the smoothing functions of the GAM model can also be estimated with assumptions, thus making the method essentially a non-parametric one. The numerical details of methods implemented in this study are omitted as this information is given exhaustively in the literature (see reference column in Table 3). The methods are implemented in R and have been tested prior to application for reference datasets.

Table 3: Applied smoothing algorithms

| | | |
|---|---|---|
| Adaptive Degree Polynomial Filter (ADP) | | |
| Auto Regressive Model (ARI) | | |
| Fast Fourier Transform Filtering (FFT) | | |
| Friedman's Super Smoother (SUP) | | |
| Generalized Additive Model (GAM) | | |
| Kalman Filtering (KAF) | | |
| Kernel Smoother (KER) | | |
| Locally-weighted polynomial (POL) | | |
| Robust Running Medians (RRM) | | |
| Savitzky-Golay Filters (SGF) | | |
| Simple Moving Average (SMA) | | |
| Spline (SPL) | | |
| Tukey Smoother (TUK) | | |
| Method | Reference | Sample |
| TUK | Mallows 1979 | Fiskeaux and Ling, 1982 |
| KAF | Welch and Bishop, 1995 | Pan et al. 2016 |
| FFT | Cochran et al. 1967 | Yang et al., 2004 |
| SPL *^ | Reinsch 1967 | Eubank 1988 |
| KER *^ | Härdle and Vieu, 1992 | Speckman, 1988 |
| SMA * | Chou, 1975 | He et al. 2020 |
| RRM * | Jerome 1982 | Polasek 1984 |
| SUP *^ | Luedicke | Friedman and Siverman, 1989 |
| POL *^ | Atkeson 1997 | Rajagopalan and Lall, 1998 |
| SGF ** | Press and Teukolsky, 1990 | Bromba and Ziegler, 1981 |
| ARI **^ | Akaike 1969 | Lohani et al. 2012 |
| ADP ***^ | Barak and Kruse | Jakubowska et al. 2004 |
| GAM****^ | Grego 2006 | Murphy et al. 2019 |

\* single parameter, ** double parameter, *** triple parameter, *** above triple parameter model and ^models with the ability of forecasting

Some of the implemented models are not only used for smoothing but also provide means for signal forecasting, e.g., GAM, ADP, ARI, POL, SUP, KER, and SPL. Since this study focuses on signal filtering, the capability and accuracy of the techniques in terms of prediction are not taken into account.

*Workflow*

The general workflow for the investigation is depicted in Figure 2. The overall data analysis is divided into two main steps a) model fitting and b) clustering. In the first step, multiple smoothing algorithms are applied to the SARS-CoV-2 titer values (raw and normalized to $NH_4$) to de-noise the measured signals in the wastewater system. To quantify their performance a cross-validation approach (Stone, 1978) is implemented to estimate a precise error value associated with each model configuration. In this context we split the SARS-CoV-2 time series $x_t$ ($x_{[1,2,...T]}$) in T subsets where a single entity x is excluded in each of the series. The T (subset) time series are denoted as $\widehat{x_t}$. For estimating the uncertainty of a smoothing model, we build and calibrate the model for each of the subsets, thus deriving T results for the performance indicators. The uncertainty of the model is subsequently estimated from this dataset.

As most models contain adjustable parameters, calibration is an essential step of the workflow. We apply mathematical optimization (as required either in discrete or continuous mode) to guarantee the best model structure/parameter selection. As stated frequently in the literature the genetic algorithm (GA) is well adapted to solve both real-valued or integer programming even for complex and ill-posed problems (e.g., Panchal and Panchal, 2015).

The second step involves clustering of the methods according to their performance (error and consistency). As some of the methods are functionally similar, the center of the clusters is seen as a representative solution. For clustering the K-Medoid algorithm (Park and Jun, 2009) is applied.

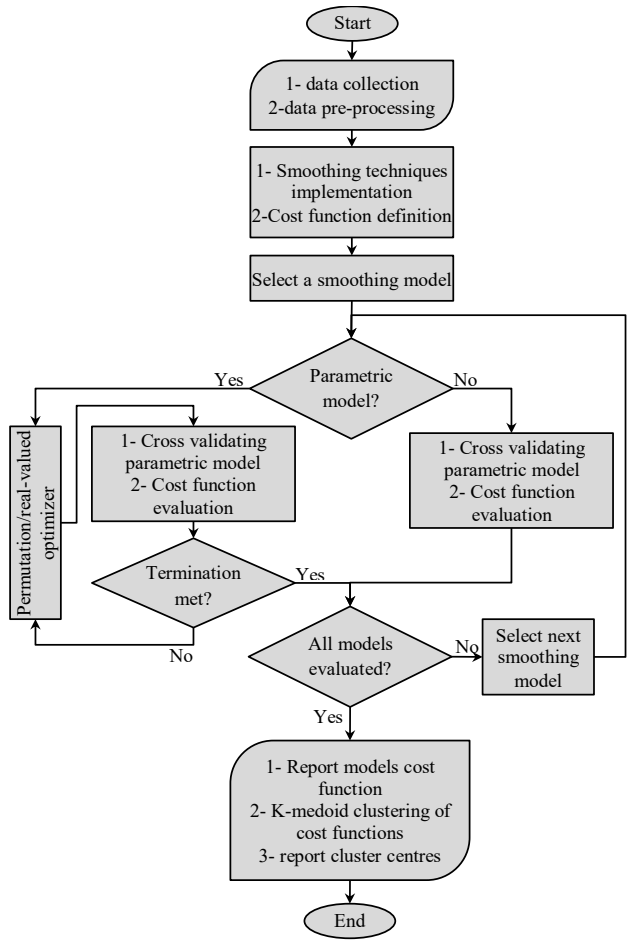

Figure 2: Flowchart for analysis of smoothing methods

*Calibration of parametric data models*

Filtering methods with parameter(s) require calibration to estimate the best configuration. Global calibration algorithms based on mathematical optimizers are manifold in science and engineering (Schutte et al. 2004; Price et al. 2006, Kaur et al. 2020). Since GA has been applied successfully to different optimization problems and is suitable to solve either permutation (integer) programming or real-valued optimization as required herein, GA was selected as optimization procedure (see Mirjalili (2019) about GA details and operators). Table 4 summarizes the parameters/configurations used for GA to calibrate the parametric filtering methods.

Table 4: GA Parameters used to calibrate filtering models

| Parameter/configuration | Value/method |
|---|---|
| Population size | 100 |
| Iteration | 1000 |
| Mutation rate | 0.1 |
| Crossover rate | 0.8 |
| Elitism | 0.05 |
| Selection | roulette wheel |
| Mutation method | Random |
| Crossover | Two points |

*Performance indices*

To measure the robustness of the filtering methods against the absence of signal entities, a multi-criteria indexing approach is performed to address the validity of methods for smoothing SARS-CoV-2 time series data. To this end, a cost function comprised of mean absolute error (MAE), variability (VAR), and Akaike Information Criterion (AIC) (Sakamoto et al. 1986) was computed for every model. With $x_t$ as the original signal and $\hat{x}$ as a column-wise square *t*-by-*t* matrix of the filtered series (where each column represents filtered values of $x_t$ under the absence of the *t*th signal value) the performance indicators MAE, VAR, and AIC are computed as:

$$MAE = \frac{\sum_t |diag\{\hat{x}\}_t - x_t|}{T} \qquad \text{Eq.4}$$

$$VAR = \sum_t \frac{\sum_t (\hat{x}_{t*} - \overline{\hat{x}_{t*}})}{T-1} \qquad \text{Eq.5}$$

$$AIC = T.\log\left(\frac{\sum_t (diag\{\hat{x}\}_t - x_t)^2}{T}\right) - 2k \qquad \text{Eq.6}$$

where $\hat{x}_{t*}$ is the *t*th row of $\hat{x}$ matrix, $\overline{\hat{x}_{t*}}$ is average of $\hat{x}_{t*}$, $k$ is the number of model parameters and $1 < t < T$.

**Results and Discussion**

Applying the methods described above, the suitability of smoothing methods is tested for the viral load signals of the 4 case studies. Note that we first apply the whole procedure as described in 2.4 to the raw signal and - in a second step - repeat the procedure for the NH-4 normalized signal.

*Raw signal*

Figure 3 shows the results of 13 implemented filtering techniques for the indicators MAE and VAR. According to the results, the MAE varies significantly from station to the station, while the indicator VAR varies similarly across stations. Generally, we can see a strong influence of the catchment size (city A-D) in the results. For the dataset City A (large city), both MAE and VAR are smaller for all filtering techniques as compared to the values obtained for the dataset City D (small community). City B and City C corroborate the trend, that MAR and VAR are increasing the smaller the catchment size. In Figure 3, the methods separated with a solid box indicate the center of the K-Medoid clustering for each station.

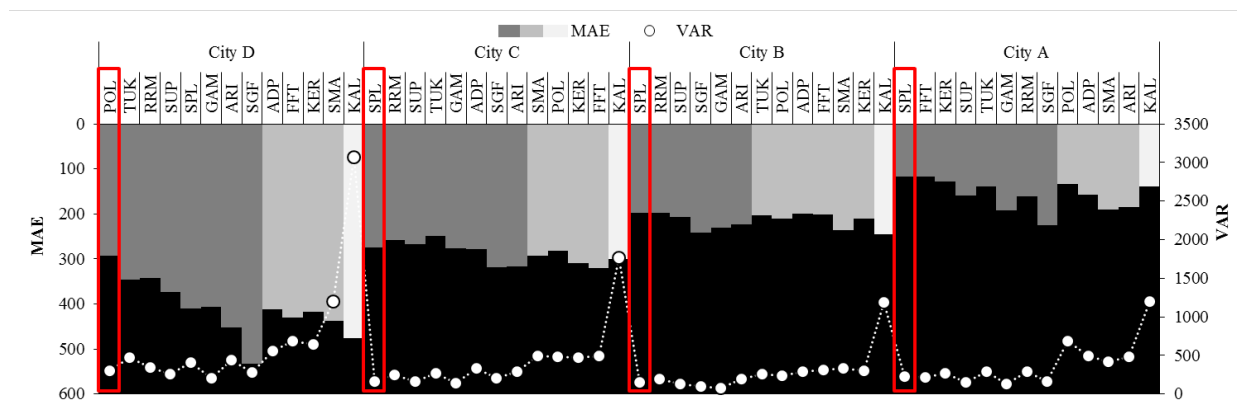

Figure 3: Performance of filtering methods for raw signals (black bars: Mean Absolute Error MAE and white dots: Variance VAR). Optimal methods are placed left for each case study.

For the K-Medoid algorithm, performance indices have been partitioned into 3 clusters namely as best, middle and worst. Next, the median of the best clusters has been defined as optimal method. As a result, for the raw signal investigation, we found the SPLINE method to exhibit best results in both MAE and VAR for the datasets A, B and C. Only in the dataset for the smallest (and touristic) catchment (City D) it is the Locally-weighted Polynomial method that behaves best. However, note that also for dataset D SPLINE is clustered among the best.

Table 5 shows the results of clustering for the filtering methods. Accordingly, the majority of the methods are placed in cluster 1, that indicates the best clustered methods. It is worth to mention that Kalman filtering (KAL) performs worst for smoothing of the viral signals in all

WWTPs. On the other hand, Friedman's Super Smoother (SUP), Spline (SPL) and Savitzky-Golay Filters (SGF) outweigh the other filtering methods in most of the cases. According to Table 5, SPL and SUP as parametric and nonparametric methods, respectively, are the only ones among the best filtering techniques in all 4 WWTPs. Two-parameter methods, ARI and SGF, are mostly clustered in the best cluster, while nonparametric methods such as TUK and FFT have been listed in both the best and the second-best clusters with the same membership frequencies.

Table 5: Clustering of performance indices for raw signal investigation

| City D | | | | City C | | | | City B | | | | City A | | | |
|---|---|---|---|---|---|---|---|---|---|---|---|---|---|---|---|---|
| Method | VAR | ERR | AIC | Method | VAR | ERR | AIC | Method | VAR | ERR | AIC | Method | VAR | ERR | AIC |
| POL | **305.9** | **292.9** | **846.6** | SPL | **159.0** | **275.3** | **662.7** | RRM | 196.5 | 198.1 | 887.1 | FFT | 220.4 | 116.2 | 723.0 |
| TUK | 475.4 | 346.5 | 860.6 | RRM | 247.6 | 258.5 | 663.8 | SPL | **154.6** | **198.3** | **889.6** | SPL | **229.8** | **117.3** | **729.9** |
| RRM | 349.4 | 343.0 | 860.9 | SUP | 167.7 | 267.0 | 664.3 | SUP | 135.3 | 207.0 | 892.8 | KER | 265.5 | 128.9 | 743.9 |
| SUP | 261.5 | 373.8 | 861.6 | TUK | 274.5 | 249.7 | 664.4 | SGF | 100.7 | 241.5 | 909.2 | SUP | 146.9 | 158.9 | 778.0 |
| SPL | 410.8 | 410.3 | 866.4 | GAM | 146.2 | 276.7 | 671.3 | GAM | 74.8 | 231.2 | 909.8 | TUK | 285.8 | 139.8 | 784.1 |
| GAM | 202.8 | 407.1 | 873.2 | ADP | 329.1 | 279.0 | 671.4 | ARI | 192.6 | 223.4 | 915.8 | GAM | 132.9 | 193.1 | 791.9 |
| ARI | 440.4 | 451.9 | 883.6 | SGF | 201.8 | 317.7 | 679.2 | TUK | 253.7 | 202.6 | 890.9 | RRM | 293.4 | 161.3 | 795.3 |
| SGF | 277.2 | 533.2 | 890.6 | ARI | 295.3 | 316.4 | 684.1 | POL | 232.4 | 210.1 | 892.0 | SGF | 158.6 | 225.9 | 814.1 |
| ADP | 560.1 | 412.6 | 871.4 | SMA | 494.3 | 293.0 | 668.3 | ADP | 295.8 | 199.5 | 895.2 | POL | 688.5 | 133.0 | 747.3 |
| FFT | 687.6 | 429.5 | 875.4 | POL | 487.5 | 281.0 | 675.2 | FFT | 309.2 | 201.4 | 905.9 | ADP | 492.6 | 157.7 | 783.3 |
| KER | 644.7 | 417.8 | 875.8 | KER | 470.6 | 309.0 | 680.3 | SMA | 333.6 | 236.6 | 909.4 | SMA | 415.8 | 189.4 | 789.8 |
| SMA | 1195.0 | 436.7 | 879.6 | FFT | 489.9 | 319.5 | 681.4 | KER | 306.1 | 210.2 | 912.9 | ARI | 478.1 | 185.2 | 803.4 |
| KAL | 3063.9 | 476.4 | 885.7 | KAL | 1762.8 | 300.6 | 672.8 | KAL | 1186.3 | 245.0 | 921.0 | KAL | 1198.8 | 138.9 | 746.4 |

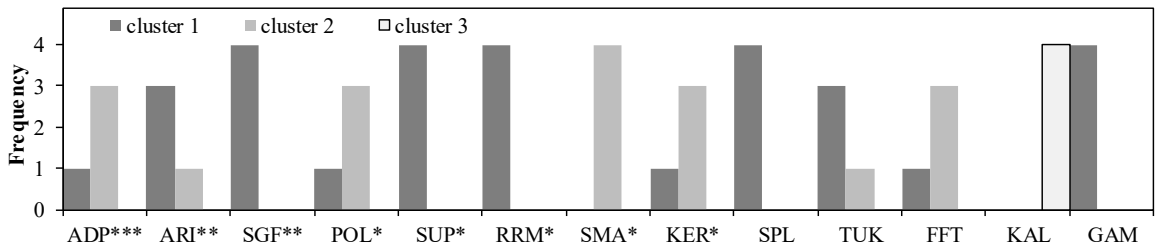

Figure 4 indicates the raw signal of viral RNA concentration measured in the four wastewater treatment plants and the smoothed signal as computed by the optimal method (center method of the best cluster). Also the result of the cross validation is depicted as well as the 95% confidence interval to assess the uncertainty of the filtering. Note that the computed uncertainty bounds are higher for City D (small population) and lower for City A (large population). Uncertainty increases where there is a peak (positive or negative) in the original data series or where a significant difference is seen between the original signal and the smoothed one.

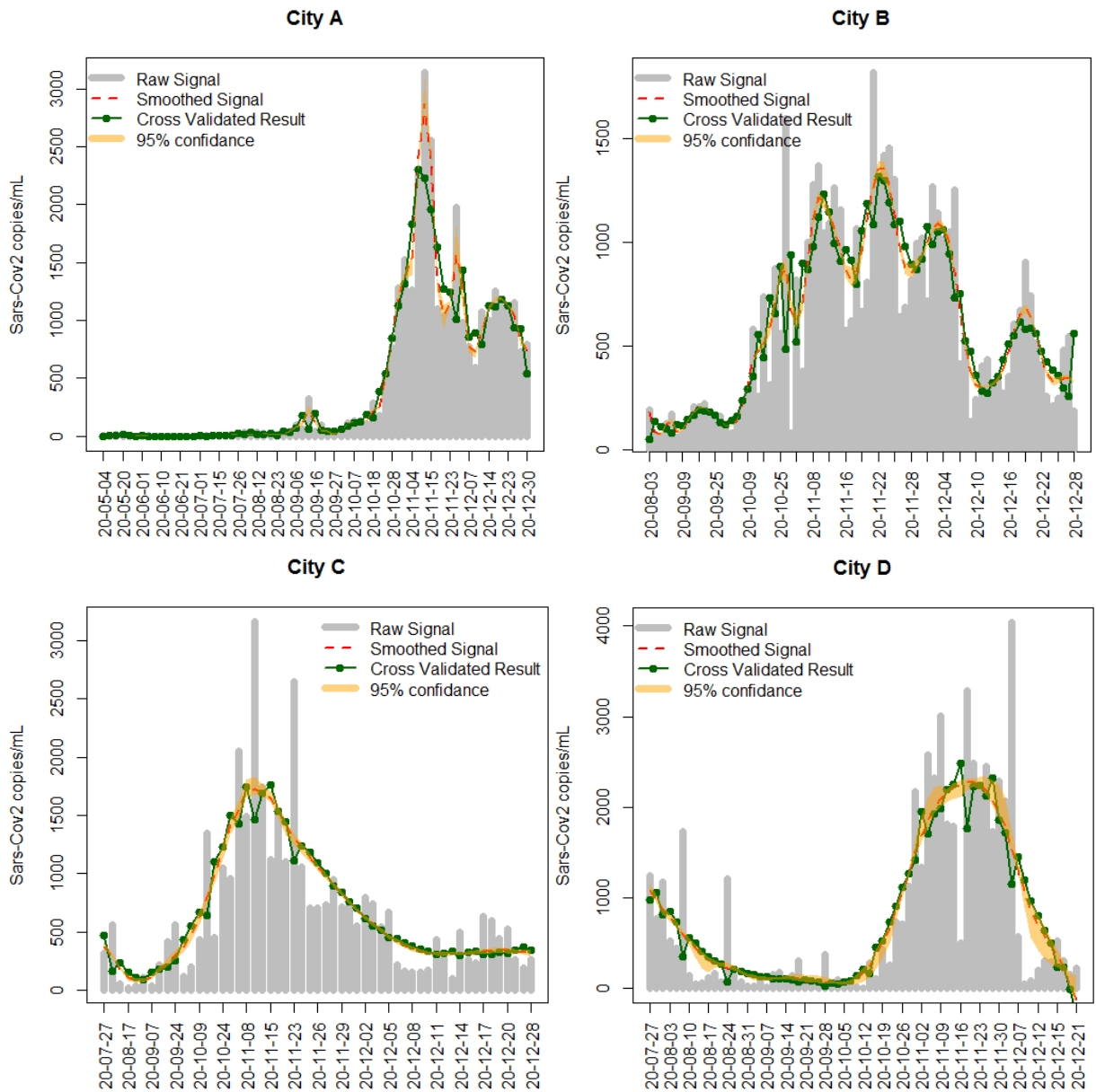

Figure 4: Application of the selected smoothing techniques to the raw data signal (superimposed with the result of cross validation and the 95% confidence interval)

*Normalized signal*

For normalizing the SARS-CoV-2 signal we apply the procedure outlined earlier by using $NH_4$ as population marker. Using the similar workflow as for the raw signal the general picture does not change drastically. For clustering the methods according to their performance, the K-Medoid algorithm was applied as well – see Table 6. As a result, POL, SPL, SUP, and GAM are identified as optimal smoothing methods. Although the optimal results (center of the best cluster) are different for the two signals (raw versus normalized), the methods SPL, GAM and

SUP are found as suitable/preferable for both signals. It is also notable that KAL again is consistently performing worst among the deployed methods.

Table 6: Clustering of performance indices for the normalized signal investigation

| City D | | | | City C | | | | City B | | | | City A | | | |
|---|---|---|---|---|---|---|---|---|---|---|---|---|---|---|---|
| Method | VAR | ERR | AIC | Method | VAR | ERR | AIC | Method | VAR | ERR | AIC | Method | VAR | ERR | AIC |
| GAM | **60.4** | **123.7** | **711.8** | SUP | **56.1** | **89.5** | **544.7** | SPL | **14.0** | **45.7** | **670.9** | POL | **69.3** | **31.8** | **550.1** |
| SGF | 78.6 | 127.1 | 707.1 | SGF | 64.0 | 96.1 | 548.6 | GAM | 14.0 | 45.6 | 680.9 | KER | 56.3 | 33.4 | 558.2 |
| SUP | 102.5 | 121.0 | 703.3 | ADP | 81.5 | 83.1 | 542.4 | SUP | 24.2 | 46.4 | 677.1 | FFT | 62.6 | 34.2 | 554.4 |
| ARI | 133.2 | 135.2 | 719.7 | TUK | 84.8 | 78.3 | 543.4 | ARI | 26.9 | 47.7 | 687.8 | SPL | 63.8 | 35.4 | 563.2 |
| SPL | 135.4 | 123.3 | 704.6 | GAM | 88.4 | 98.8 | 564.9 | SGF | 29.1 | 47.2 | 680.3 | KAL | 261.6 | 40.6 | 581.4 |
| RRM | 142.4 | 108.3 | 699.3 | RRM | 88.9 | 79.1 | 544.5 | RRM | 40.7 | 44.6 | 677.0 | TUK | 85.0 | 41.8 | 611.1 |
| SMA | 143.2 | 111.0 | 694.8 | SMA | 125.3 | 97.6 | 560.2 | ADP | 52.1 | 47.2 | 685.5 | SMA | 80.0 | 43.0 | 593.9 |
| ADP | 161.5 | 125.4 | 706.8 | ARI | 144.6 | 120.4 | 576.2 | TUK | 55.4 | 46.4 | 681.8 | RRM | 105.0 | 44.5 | 613.0 |
| TUK | 163.1 | 112.7 | 704.7 | KER | 164.6 | 107.0 | 566.7 | SMA | 63.9 | 49.5 | 683.9 | SUP | 37.0 | 44.7 | 596.7 |
| KER | 204.5 | 127.6 | 719.6 | SPL | 196.5 | 109.7 | 568.5 | KER | 83.4 | 53.8 | 707.4 | SGF | 42.6 | 47.0 | 603.2 |
| FFT | 205.6 | 124.7 | 713.7 | FFT | 218.2 | 117.5 | 592.4 | FFT | 83.6 | 50.8 | 699.3 | ADP | 91.7 | 50.0 | 618.1 |
| POL | 247.7 | 117.5 | 713.3 | POL | 218.2 | 117.5 | 594.4 | POL | 83.6 | 50.8 | 701.3 | GAM | 34.2 | 51.3 | 611.0 |
| KAL | 723.0 | 129.1 | 703.1 | KAL | 481.3 | 106.1 | 559.7 | KAL | 203.0 | 49.1 | 687.8 | ARI | 108.3 | 51.8 | 630.4 |

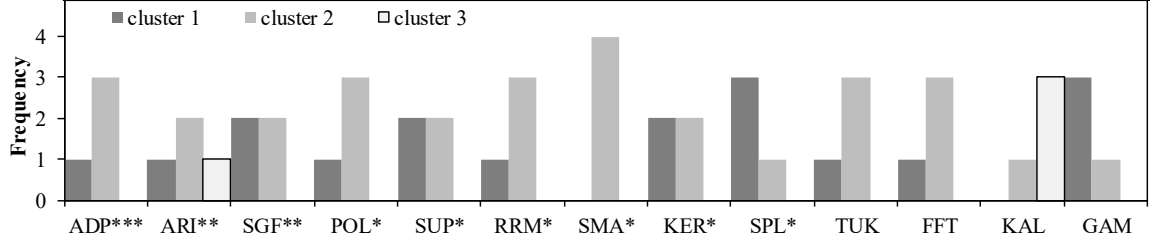

Similar to above, all normalized time series are depicted by using the optimal signal filtering method based on the clustering (Figure 5). Despite the difference in scale, the shape of the smoothed data is quite similar for both normalized and raw signals. A difference is the higher variation of the cross-validation results – indicating a higher sensitivity of the selected filtering techniques when applied to the normalized signal as compared to the raw data.

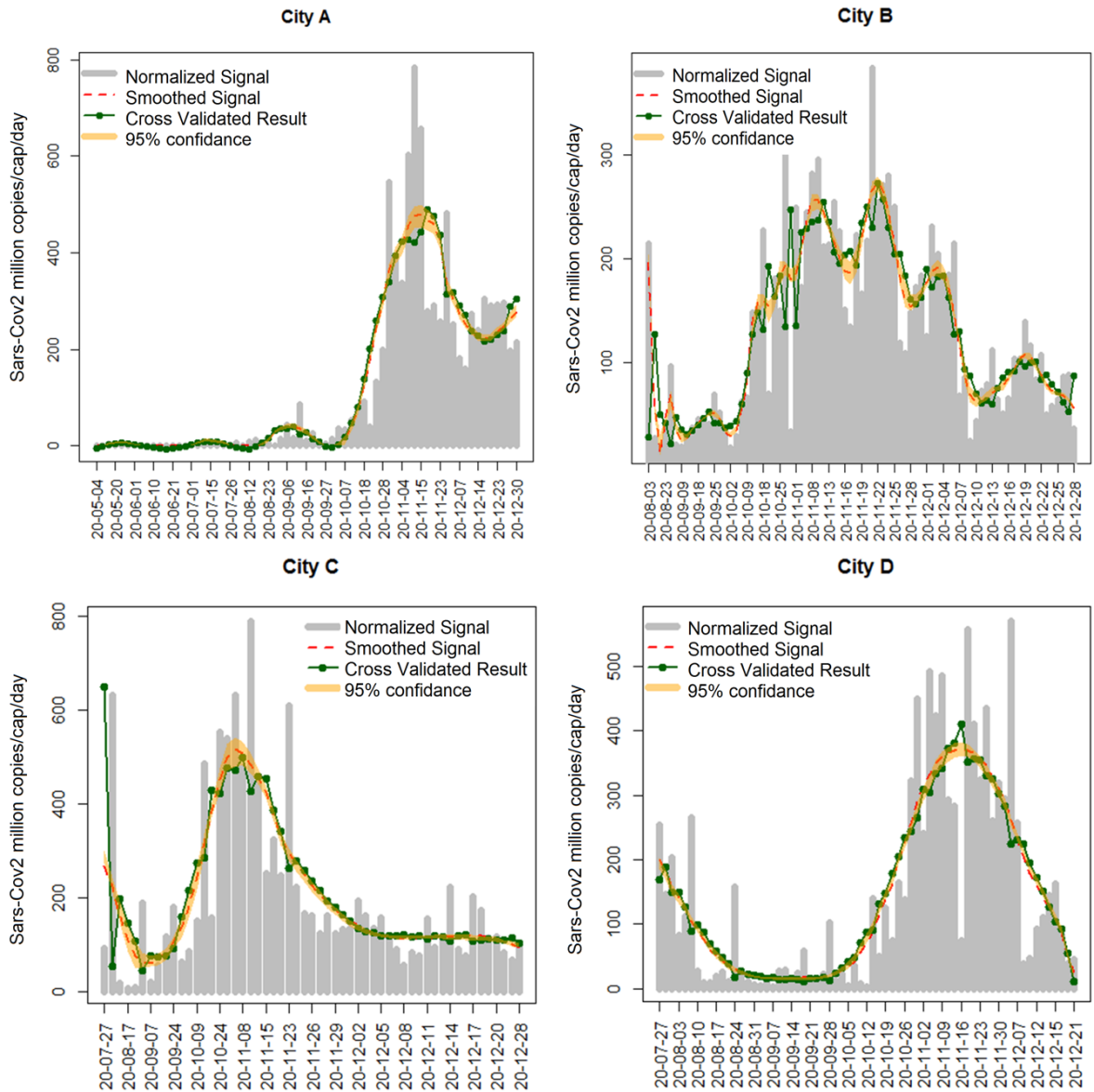

Figure 5: Application of the selected smoothing techniques to the normalized data signal (expressed as RNA $10^6$ copies/cap/day) superimposed with the result of cross validation and the 95% confidence interval

*Virological Analysis*

Additional to the analysis of the smoothing methods, the dataset also allows to reflect on the relation of the wastewater measurements to the infection dynamics. As a ballpark approximation we plot the ratio of the viral load per capita to the 7-day incidence value (expressed as weekly cases per 100,000 persons) by fitting linear models to each case study (Figure 6). Using the marginal probability densities, we derive for the 50-percentile value of the viral load (appr. $100 \times 10^6$ copies/cap/day) incidence values ranging between 100 – 600 (7

day/100,000 persons). The interesting feature is the influence of scale: From the linear models we see both higher intercepts and higher slopes for small catchments and vice versa. The first observation indicates that small catchments have a certain threshold of viral load before a clear relation with infection dynamics is seen. The second observation (i.e., the slope of the model) points to the fact that infection clusters are (statistically) more significant in a smaller population than in a large one. Overall, community size seems to be an influential factor for WBE in the case of SARS-Cov-2.

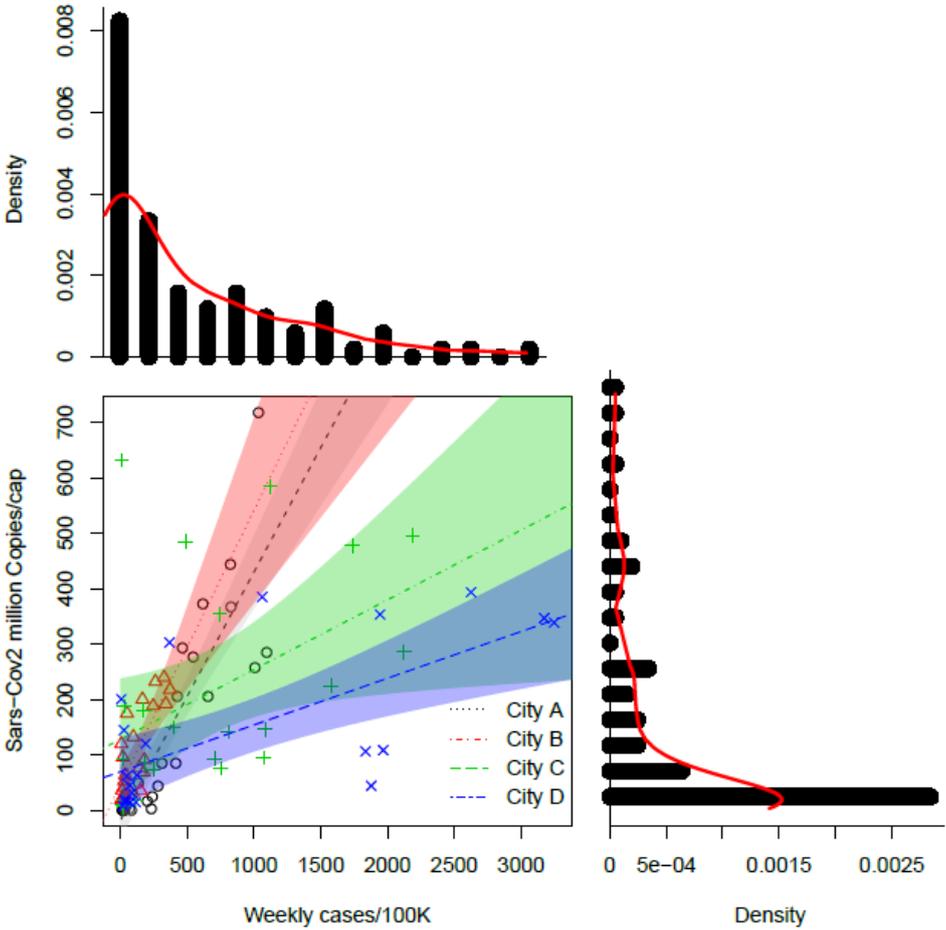

Figure 6: Linear models expressing the ratio of normalized viral loads against 7-day incidence per 100 000 persons.

**Conclusion**

For management of pandemics such SARS-CoV-2 and interpretation of data obtained by means of WBE, filtering of the wastewater titer signal is an important pre-processing step. Modeling the infection dynamics or developing predictive tools therefrom may induce misleading results when based on noisy information. This is especially important when models are not robust enough against extreme/oscillative inputs. This study focused on the application of 13 well-established signal filtering techniques for smoothing of SARS-CoV-2 datasets in four wastewater treatment plants across Austria. Based on the finding in this study the following conclusions are made:

- Spline, GAM and Friedman's Super Smoother are recognized as superior methods in this context. In three wastewater treatment plants Spline was found as robust approach to cope with missing data and uncertainties.
- Although GAM is a robust smoother against extremes and outliers, it either requires a high number of parameters to be tuned, or – applying it as non-parametric method - tends to over-smooth signals. The latter also applies to the Friedman's Super Smoother technique.
- For the case of nonparametric methods, TUK and FFT performed generally well and are suitable algorithms. However, nonparametric methods are sensible for missing values and are thus only recommended for times series with a small number of missing signals.
- Despite acceptable error values for methods such as KAL, SMA, and POL, they are not suitable in this context as potentially overfitting.
- A first analysis of the dataset indicates that community size has an influence on WBE for SARS-Cov-2. For smaller catchments both a threshold of viral load is apparent before any relation with infection dynamics is visible and also a higher sensitivity towards infection clusters.

Overall, all filtering methods fit the general purpose, that is data smoothing. What constitutes a superior method is depending on the aim, the dataset and the context. For this investigation i.e., smoothing the SARS-CoV-2 signal and applying the metrics: minimizing MAR and VAR – we advocate to use the SPLINE method as it gives the best balance of result versus robustness or GAM if a non-parametric method is the aim.

**Acknowledgements**

This study was funded by the Austrian Federal Ministry of Education, Science and Research, the Austrian Federal Ministry of Agriculture, Regions and Tourism, the federal states Burgenland, Carinthia, Lower Austria, Salzburg, Styria, Tirol, Upper Austria and Vorarlberg and the Austrian Association of Cities and Towns. We would like to thank the staff of the treatment plants involved for their support and the use of the wastewater titer measurement data. The support of Günther Weichlinger and Christoph Scheffknecht are also appreciated.